\documentclass[sigconf]{acmart}
\AtBeginDocument{%
  }

\copyrightyear{2025}
\acmYear{2025}
\setcopyright{acmlicensed}
\acmConference[MM '25]{Proceedings of the 33rd ACM International Conference on Multimedia}{October 27--31, 2025}{Dublin, Ireland}
\acmBooktitle{Proceedings of the 33rd ACM International Conference on Multimedia (MM '25), October 27--31, 2025, Dublin, Ireland}
\acmDOI{10.1145/3746027.3755758}
\acmISBN{979-8-4007-2035-2/2025/10}

\usepackage{multirow}
\usepackage{bbding,pifont,utfsym}
\usepackage{enumitem}
\begin{document}

\title{AudioGenie: A Training-Free Multi-Agent Framework for Diverse Multimodality-to-Multiaudio Generation}

\author{Yan Rong}
\affiliation{%
  \institution{The Hong Kong University of Science and Technology (Guangzhou)}
  \city{Guangzhou}
  \state{}
  \country{China}
}
\email{yrong854@connect.hkust-gz.edu.cn}

\author{Jinting Wang}
\affiliation{%
  \institution{The Hong Kong University of Science and Technology (Guangzhou)}
  \city{Guangzhou}
  \state{}
  \country{China}
}
\email{jwang644@connect.hkust-gz.edu.cn}

\author{Guangzhi Lei}
\affiliation{%
  \institution{Tencent AI Lab}
  \city{Shenzhen}
  \state{}
  \country{China}
}
\email{shaanyang@tencent.com}

\author{Shan Yang}
\affiliation{%
  \institution{Tencent AI Lab}
  \city{Shenzhen}
  \state{}
  \country{China}
}
\email{guangzhilei@tencent.com}

\author{Li Liu}\authornote{Corresponding Author.}
\affiliation{%
  \institution{The Hong Kong University of Science and Technology (Guangzhou)}
  \city{Guangzhou}
  \state{}
  \country{China}
}
\email{avrillliu@hkust-gz.edu.cn}
\renewcommand{\shortauthors}{Rong et al.}


\begin{abstract}
Multimodality-to-Multiaudio (MM2MA) generation faces significant challenges in synthesizing diverse and contextually aligned audio types (\textit{e.g.}, sound effects, speech, music, and songs) from multimodal inputs (\textit{e.g.}, video, text, images), owing to the scarcity of high-quality paired datasets and the lack of robust multi-task learning frameworks. Recently, multi-agent system shows great potential in tackling the above issues. However, directly applying it to MM2MA task presents three critical challenges: (1) inadequate fine-grained understanding of multimodal inputs (especially for video), (2) the inability of single models to handle diverse audio events, and (3) the absence of self-correction mechanisms for reliable outputs. 
To this end, we propose \textbf{AudioGenie}, a novel training-free multi-agent system featuring a dual-layer architecture with a generation team and a supervisor team. For the generation team, a fine-grained task decomposition and an adaptive Mixture-of-Experts (MoE) collaborative entity are designed for detailed comprehensive multimodal understanding and dynamic model selection, and a trial-and-error iterative refinement module is designed for self-correction. The supervisor team ensures temporal-spatial consistency and verifies outputs through feedback loops. Moreover, we build MA-Bench, the first benchmark for MM2MA tasks, comprising 198 annotated videos with multi-type audios. 
Experiments demonstrate that our AudioGenie achieves state-of-the-art (SOTA) or comparable performance across 9 metrics in 8 tasks. User study further validates the effectiveness of our method in terms of quality, accuracy, alignment, and aesthetic.
The project website with audio samples can be found at \url{https://audiogenie.github.io/}.
\end{abstract}






\maketitle

\section{Introduction}
\begin{table*}[thb]
\centering
\caption{Comparison with state-of-the-art methods across various aspects (inputs, outputs and the main characters).}
\resizebox{\linewidth}{!}{
    \begin{tabular}{c|cccc|ccccc|cc}
    \toprule
    \multirow{2}{*}{Methods} & \multicolumn{4}{c|}{Inputs} & \multicolumn{5}{c|}{Outputs} & \multicolumn{2}{c}{Characters}\\
    \cmidrule(r){2-5} \cmidrule(r){6-10} \cmidrule(r){11-12}
    & Text & Image & Video& Combination & Sound Effects & Speech & Song & Music & Combination & Agent-based &  Self-Correct \\
    \midrule
    Auffusion~\cite{xue2024auffusion}& \ding{52} & \ding{56}& \ding{56}& \ding{56} & \ding{52} & \ding{56}& \ding{56}& \ding{56}& \ding{56}& \ding{56}& \ding{56} \\
    % VidMuse~\cite{tian2024vidmuse}&\ding{56} &\ding{56} & \ding{52} & \ding{56}&\ding{56} &\ding{56} &\ding{56} & \ding{52} & \ding{56}&\ding{56} &\ding{56}\\
    MMAudio~\cite{cheng2024taming}& \ding{52} & \ding{56} & \ding{52}& \ding{52} & \ding{52}& \ding{56}&\ding{56} & \ding{56}& \ding{56} & \ding{56} & \ding{56} \\
    FoleyCrafter~\cite{zhang2024foleycrafter}& \ding{52} & \ding{56} & \ding{52}& \ding{52} & \ding{52}& \ding{56}&\ding{56} & \ding{56}& \ding{56} & \ding{56} & \ding{56} \\
    AudioX~\cite{tian2025audiox}&\ding{52} &\ding{52} &\ding{52} &\ding{52}& \ding{52}&\ding{56} & \ding{56}&\ding{52} & \ding{52}&\ding{56}&\ding{56}\\
    \midrule
    LVAS-Agent~\cite{zhang2025long} &\ding{56} &\ding{56} & \ding{52} & \ding{56}& \ding{52} &\ding{56} & \ding{56}& \ding{56}& \ding{56}&  \ding{52} &\ding{56} \\
    \midrule
    AudioGenie (Ours) &\ding{52}& \ding{52} & \ding{52} & \ding{52} &\ding{52}&\ding{52} & \ding{52}&\ding{52} & \ding{52}& \ding{52}& \ding{52} \\
    \bottomrule
    \end{tabular}
}
\label{intro_domains}
\end{table*}

Multimodality-to-Multiaudio (MM2MA) focuses on generating multiple audio types (\textit{i.e.}, sound effects, speech, song, music, or their combinations) from diverse multimodal inputs (\textit{i.e.}, video, text, image, or a mix of them). By enabling varied and context-tailored audio outputs, MM2MA opens up new opportunities for multimodal human-computer interactions, such as game development~\cite{friberg2004audio}, film production~\cite{cong2024styledubber}, and VR/AR experiences~\cite{yang2022audio}.

In the literature, great progress in text-to-audio (T2A)~\cite{xue2024auffusion,huang2023make,majumder2024tango} generation and video-to-audio (V2A)~\cite{cheng2024taming,zhang2024foleycrafter,xing2024seeing} has been achieved by prior works. However, as shown in Table~\ref{intro_domains}, the scarcity of large-scale paired datasets still limits the development of a unified MM2MA framework, which can handle diverse inputs and produce various types of audio. One representative effort in this direction is AudioX~\cite{tian2025audiox}, recently posted on arXiv, which employs a diffusion-based Transformer to generate audio from multimodal inputs. However, it focuses mainly on audio effects and music (no speech and song), and its training process remains computationally intensive. Moreover, the paired dataset required for training has not been made public. Therefore, a more all-encompassing, data-friendly, and open-source MM2MA generation approach is increasingly desirable.

Recently, autoagent-based paradigms have shown potential in handling multi-task~\cite{chen2024autoagents,huang2024audiogpt,liang2024wavcraft} and data-scarce scenarios~\cite{li2024anim,tu2024spagent}.
However, directly applying the vanilla autoagent-based paradigm to MM2MA generation suffers from three main challenges:
\textbf{(1) Input Understanding Issue}: Limited ability of understanding fine-grained multimodal inputs. Multimodal inputs, especially video, require fine-grained understanding and temporal-spatial consistency capturing, which existing vanilla agent methods struggle to handle.
\textbf{(2) Selecting Issue}: A single agent or model cannot effectively manage all types of audio events. Each audio type requires domain-specific expertise. Moreover, even within the same task, different models often posses different strengths and focus. It is nearly impossible for one well-trained model to satisfy all user needs, and relying on a single model is typically suboptimal.
\textbf{(3) Self-Correcting Issue}: Models lack the ability to automatically self-correct results, leading to unreliable outputs. Existing pipelines typically generate audio in a single pass without iterative feedback loops, making it hard to fix errors or integrate late-stage adjustments. Even well-trained models may misalign audio events with user inputs or produce artifacts that remain unchecked. 
In fact, some agent-based methods (\textit{e.g.}, ~\cite{zhang2025long}) have been explored for single-input single-output task (\textit{e.g.}, video to sound-effect generation), yet they still face challenges mentioned above.

Therefore, in this work, the main question we aim to address is: \textit{How can we develop a unified autoagent-based audio generation system that supports a broader range of audio events to meet diverse human needs while addressing the three challenges mentioned above?}

To address the above three challenges, we propose a novel unified training-free multi-agent system, named \textbf{AudioGenie}, which features a dual-layer coordinated structure of a generation team and a supervisor team. The generation team operates in three stages, each targeting one of the aforementioned challenges. The supervisor team plays the different roles to monitor and correct each step in the pipeline, including detecting inaccuracies, prompting self-reflection, and driving iterative corrections. 

More precisely, \textbf{for the first challenge}, guided by the supervisor team, we introduce a fine-grained task decomposition module, which systematically breaks down the overall audio requirement into manageable sub-events, ensuring each sub-task is managed with clarity and precision. 
\textbf{For the second challenge}, we develop an adaptive Mixture-of-Experts (MoE) collaborative entity, featuring two novel complementary mechanisms: intra-expert self-refinement and inter-expert collaborative refinement. This design fosters knowledge sharing and continual performance gains. Besides, each expert is specialized for a particular audio domain to adaptively pick the most suitable model based on prior knowledge and the current context and dynamically provide fine-grained refinement for each audio event, avoiding a one-size-fits-all approach. 
\textbf{For the third challenge}, we propose a trial-and-error iterative refinement module that lets the agent integrate feedback for incremental improvement. For each audio event, the agent constructs a tree-of-thought (ToT) and employs self-correction and feedback to dynamically evaluate and revise outputs. Through this “try–reflect–optimize” loop, the proceeding of the single generation can be considered as a traversal process.

In addition, to enable comprehensive evaluation, we present the \textbf{M}ulti-type \textbf{A}udio synthesis \textbf{Bench}mark (\textbf{MA-Bench}), which comprises 198 manually curated videos spanning diverse domains. Each video includes our detailed annotation (\textit{e.g.}, audio type, timestamps for different audio events, descriptions), multi-type generated audio, and ground-truth audio, serving as a valuable resource for the MM2MA research.

In summary, the main contributions of this work are as follows: 
\begin{itemize}
\item A unified training-free multi-agent system, named AudioGenie, consisting of a generation team and a supervisor team, is proposed. To the best of our knowledge, this is the first attempt to cover such a comprehensive range of input modalities and audio types in a unified framework. 
\item We design a fine-grained task decomposition module that breaks down complex audio requirements into manageable sub-events for more detailed understanding. Also, we introduce an adaptive MoE collaborative entity incorporating intra-expert self-refinement and inter-expert collaborative refinement for adaptive model selection and dynamic refinement of audio event planning.
\item We present a novel trial-and-error iterative refinement module that evaluates and revises generation results by constructing a tree-of-thought for each audio event, enabling self-correction and incremental improvements.
\item We build the first MM2MA benchmark called MA-Bench. Objective and subjective experimental results on our proposed MA-Bench show that our AudioGenie outperforming state-of-the-art (SOTA) methods on various tasks. 
\end{itemize}

\section{Related Work}
\begin{figure*}[t]
    \centering
    \includegraphics[width=0.99\textwidth]{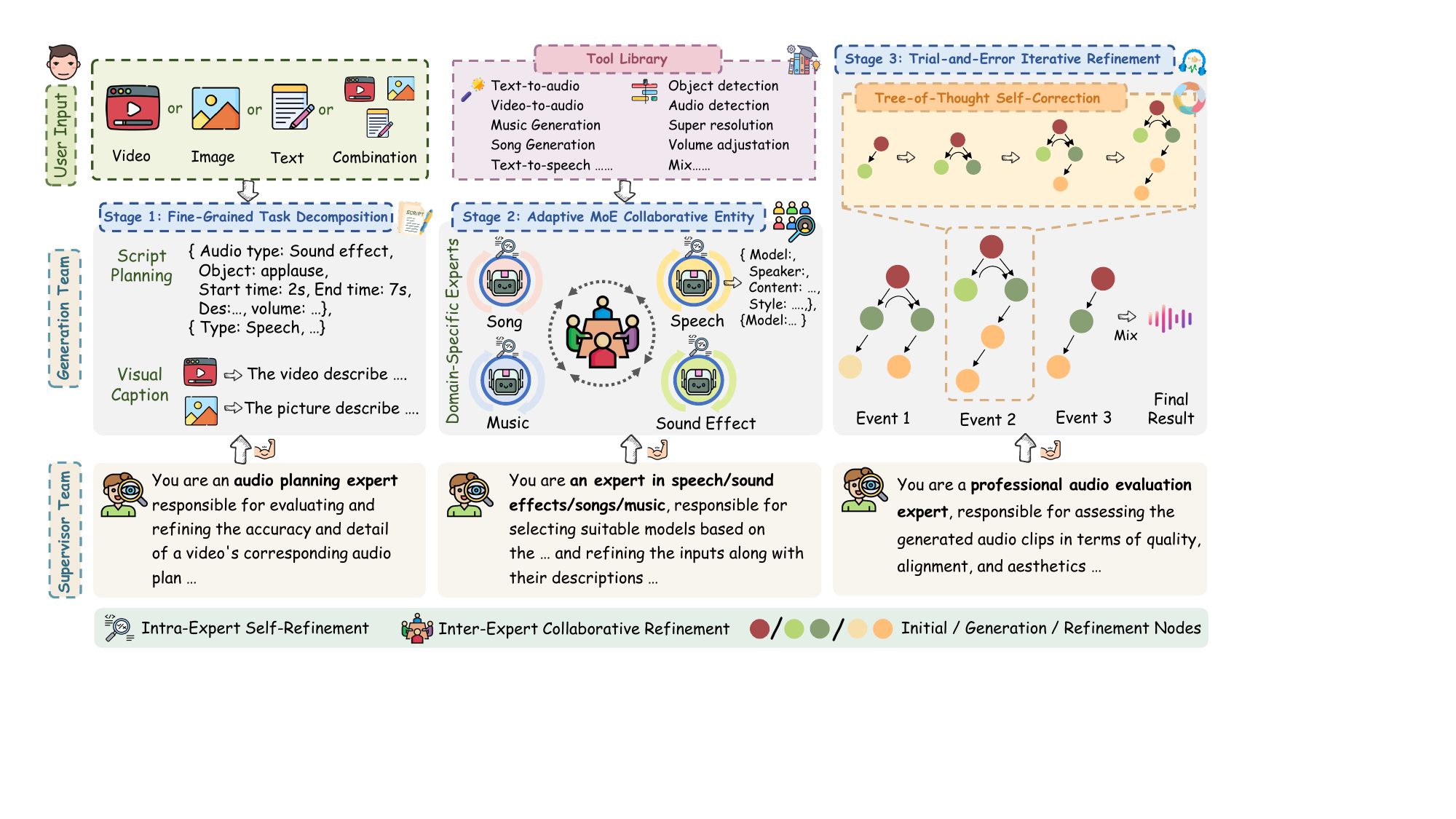}
    \vspace{-0.2em}
    \caption{Overview of the proposed AudioGenie, which supports various input and multiple types of audio outputs.}
    \vspace{-0.2em}
    \label{fig:framework}
\end{figure*}

\subsection{ X-to-Audio Generation}
\textit{X-to-audio} generation, which aims to synthesize various types of audio such as music, sound effects, speech, and song from various inputs (\textit{i.e.}, $X$ modality can be video, text, images or any combination), is an emerging area of research that faces significant challenges in multi-modal understanding and generation.

Current audio generation models typically target a single input modality or a limited conditioning type. For instance, \textit{text-to-audio} methods generate speech \cite{du2024cosyvoice2, guo2024fireredtts, xie2024towards, rong2025seeing, rong2025dopamine}, sound effects \cite{yuan2024retrieval, huang2023make, liu2023audioldm, xue2024auffusion}, music \cite{copet2023simple, chen2024musicldm, melechovsky2023mustango, liu2024music}, or songs \cite{lei2024songcreator, hong2024text, ding2024songcomposer, liu2025songgen} from textual descriptions.
\textit{Video-to-audio} approaches produce speech \cite{lei2024uni, choi2025v2sflow, gupta2024visatronic, kim2025faces}, music \cite{tian2024vidmuse, li2024muvi, xie2025filmcomposer, zuo2025gvmgen,wang2025motioncomposer}, or sound effects \cite{cheng2024taming, wang2024v2a, comunita2024syncfusion, ren2025sta, zhang2025long} synchronized with visual cues. 
While these specialized models excel in their respective tasks, they lack flexibility for more complex multimodal scenarios.

The demand for holistic audio synthesis has driven the development of unified frameworks capable of generating multiple audio types under multimodal conditions. For instance, AudioX \cite{tian2025audiox} presents a single model that flexibly handles text, video, and image inputs to produce both sound effects and music. However, such models are data-hungry, demanding large-scale paired datasets and costly training procedures.

Instead of relying on extensive training, we propose a training-free unified multi-agent system for efficient, collaborative generation of multiple audio types from multimodal inputs.

\vspace{-0.2em}
\subsection{Auto-agent System}

Several studies utilize multimodal large language models (MLLMs) as core AI agents, enabling intelligent planning and decision-making to address more complex tasks, such as image generation and editing~\cite{wang2024genartist}, video generation~\cite{tu2024spagent,li2024anim}, and system development~\cite{qian2023communicative}.

Although MLLM agents have been applied to many tasks, their use in audio generation remains limited. Early systems like AudioGPT~\cite{huang2024audiogpt} and WavCraft~\cite{liang2024wavcraft} use an agent to coordinate multiple expert audio models for basic understanding and generation. More recently, LVAS-Agent~\cite{zhang2025long} proposes a multi-agent framework for long-video audio synthesis. However, this method focuses only on sound effects.

Unlike these approaches, our method proposes a unified, training-free multi-agent system specifically designed for multi-type audio generation from multimodal inputs, capable of meeting diverse user needs while ensuring reliable outputs.

\section{Method}
The overview of our proposed AudioGenie is illustrated in Figure~\ref{fig:framework}. The key idea is to cultivate a multimodal large language model (MLLM) to serves as the “brain”, coordinating and correcting the entire generation process across different audio types. The system operates under a dual-layer coordinated structure with three key novel components: the Fine-Grained Task Decomposition Module, the Adaptive MoE Collaborative Entity, and the Trial-and-Error Iterative Refinement Module.

\subsection{AudioGenie: Dual-Layer Coordinated Structure}
Single-pass generation methods often struggle to produce high-quality audio for complex, multi-event scenarios. To address this, we introduce AudioGenie with a dual-layer coordinated structure, composed of a generation team and a supervisor team.

\textbf{Generation Team.}
This component focuses on analyzing and generating audio content through three hierarchical stages. Firstly, it breaks down the complex audio requirements into manageable sub-events for fine-grained understanding. Secondly, it assigns each sub-event to an appropriate domain-specific expert for further model selection and refinement. Finally, the generation team performs the audio synthesis for each sub-event, supported by a tree-of-thought iterative refinement process.

\textbf{Supervisor Team.}
Prior work has shown that assigning complex reasoning tasks to specialized roles can improve both efficiency and quality~\cite{zhangautomatic}. Considering this, the supervisor team monitors and provides feedback at each stage through a dynamic domain-specific expert simulation way, where supervisor is role-conditioned to focus on a specific aspect. Specifically, in the first stage, the supervisor acts as an audio planning expert to evaluate the plan from content suitability, timing accuracy, and alignment, and provide concrete suggestions and revise them if necessary. In the second stage, the supervisor acts as a domain expert to assess model choices and refined JSON files for each audio event type. In the third stage, supervisor serves as a professional audio evaluator, collaborating with the generation team to perform on-the-fly self-corrections.

This coordinated “plan-and-verify” paradigm fosters effective collaboration among multiple agents, maintaining flexibility while ensuring that final outputs remain aligned with input requirements.

\subsection{Fine-Grained Task Decomposition Module}
Multimodal inputs often contain multiple overlapping or interacting long-duration audio events. To achieve fine-grained coordination across time, content, and style for these events, we design a fine-grained task decomposition module that splits the overall audio requirements into a series of manageable sub-audio events. Specifically, the agent analyzes the semantic and temporal information in the input to identify implied audio events. Each aduio event is assigned an audio type (\textit{i.e.,} speech, sound effect, song, or music), a description, and a timestamp. The resulting plan is represented as a structured JSON list with content \textit{\{audio\_type, object, start\_time, end\_time, description, volume\}}.  For any included visual information, the agent is prompted to generate captions that describe the overall scene, providing a unified ambience for subsequent style planning.

\begin{table}[t]
\centering
\caption{Models in Tool Library.}
\vspace{-0.2em}
% \resizebox{\linewidth}{!}{
    \begin{tabular}{c|c}
    \toprule
    % \multicolumn{2}{c|}{Generation Model} & \multicolumn{2}{c}{Auxiliary tools}\\
    Tasks & Models \\
    \midrule
    Video/Text-to-Sound Effect Generation& MMAudio~\cite{cheng2024taming} \\   
    % Text-to-Sound Effect Generation& MMAudio~\cite{cheng2024taming} \\
    Text-to-Sound Effect Generation& Auffusion~\cite{xue2024auffusion} \\
    Text-to-Speech Generation& CosyVoice 2~\cite{du2024cosyvoice2} \\
    Text-to-Speech Generation& FireRedTTS~\cite{guo2024fireredtts}\\
    Text-to-Music Generation& InspireMusic~\cite{zhang2025inspiremusic} \\
    Text-to-Music Generation& MusicGen~\cite{copet2023simple} \\
    Lyrics-to-Song Generation & DiffRhythm~\cite{ning2025diffrhythm}\\
    Super Resolution & AudioSR~\cite{liu2024audiosr}\\
    Audio Extraction & AudioSep~\cite{liu2024separate} \\   
    \bottomrule
    \end{tabular}
% }
\vspace{-0.2em}
\label{tool_library}
\end{table}

\subsection{Adaptive MoE Collaborative Entity}
Different audio types have distinct characteristics, making it difficult for a single agent to handle all details effectively. The success of MoE in language models demonstrates that specialized, fine-grained operations can significantly boost performance~\cite{shazeer2017outrageously}. To address this challenge, we propose an Adaptive MoE Collaborative Entity, where each audio type is managed by a dedicated domain expert rather than a shared generalist agent. 

Specifically, in the second stage, the agent assigns each audio event that generated in the first stage to the corresponding expert based on its type. Each expert then selects the suitable generation models from our tool library, considering both the audio type and the planning information in the JSON file. The tool library includes a collection of SOTA models tailored for different audio types. Notably, due to varying training datasets and architectures, different models emphasize different aspects and exhibit strengths in specific themes or scenarios. Therefore, even for the same task, we include multiple model options, as shown in Table~\ref{tool_library}. The tasks, input formats, and specific characteristics of each model are provided as prior knowledge to each domain expert. For each audio event, the expert selects two candidate models and refines the corresponding JSON file according to the input specifications of the chosen models. This approach harnesses the complementary strengths of multiple experts and models to address complex tasks effectively.

\textbf{Intra-expert Self-Refinement.}
To detect potential biases within each expert, we introduce a self-reflection loop in which the expert examines its own plan. Through this self-critique thinking mode, the expert identifies potential flaws, omissions, or ambiguities and iteratively refines the plan based on its specialized knowledge. This mechanism allows each expert to reflect on and improve its own output, increasing the accuracy of sub-event planning.

\textbf{Inter-expert Collaborative Refinement.}
To further encourage knowledge sharing, the agent will enter into inter-expert collaborative refinement phase guided by the feedback provided by the supervisor team. In this phase, experts refine the collective plan in a predefined order, offering insights from their specialized perspectives. This collaborative process allows the expert to utilize diverse expertise, covering blind spots that any single expert might miss. Such a group discussion aims to reach a consensus plan superior to any single agent’s initial proposal, as each expert’s strengths can compensate for others’ weaknesses.

\begin{figure*}[t]
    \centering
    \includegraphics[width=0.93\linewidth]{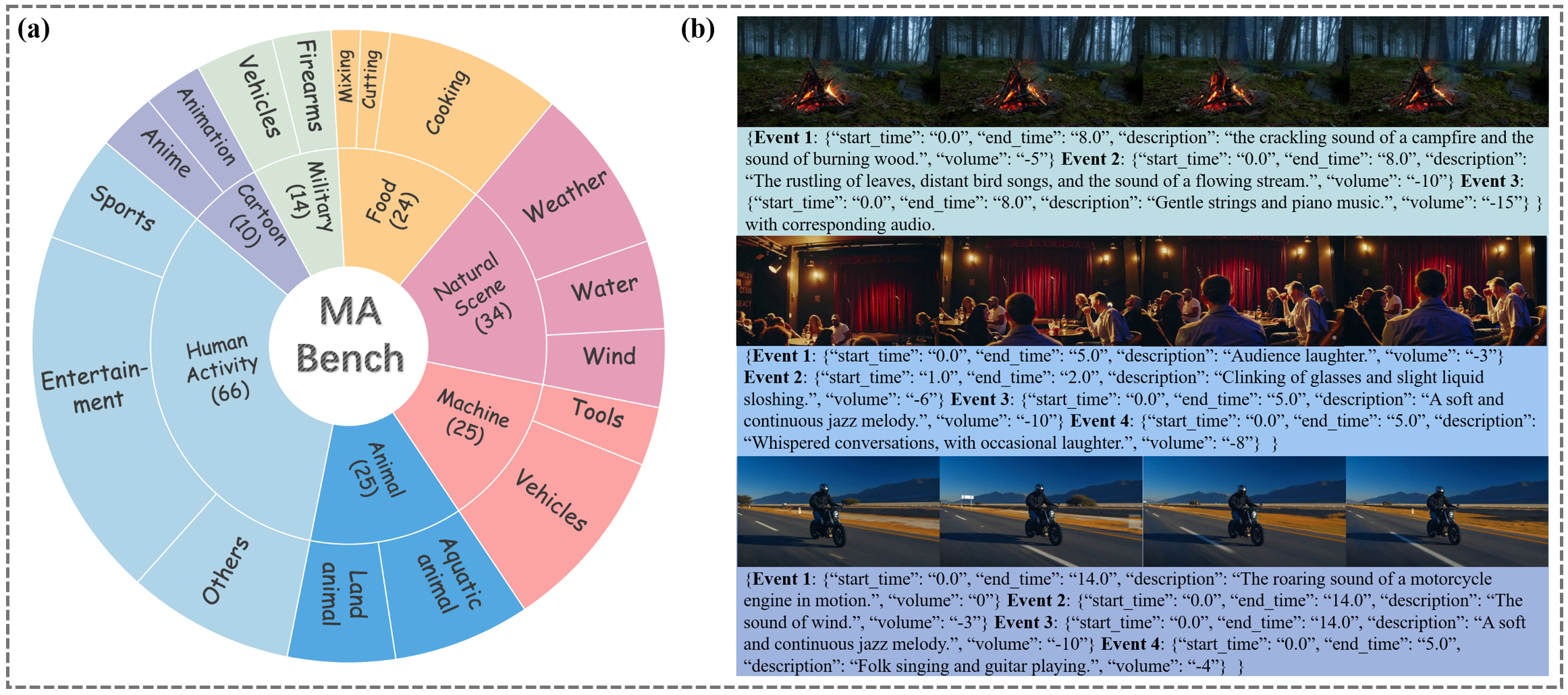}
    \caption{Statistics of video categories within our \textbf{MABench}. (a) provides statistical
distributions of video categories and sub-categories across the dataset. (b) illustrates sample data from three data sources in the MABench.}
    \label{fig:dataset}
\end{figure*}

\subsection{Trial-and-Error Iterative Refinement Module}

Single-pass audio generation often suffers from unstable output quality and poor coherence. Inspired by ~\cite{wang2024genartist}, we propose the Trial-and-Error Iterative Refinement Module. We organize the generation process of each audio event as a tree-of-thought structure, enabling the agent to explore multiple generation paths and adjust or backtrack as needed. This structure consists of three types of nodes: an initial node, generation nodes, and refinement nodes. The initial node is the root of the tree and receives the corresponding JSON file, marking the beginning of the generation. The generation node produces the audio event based on the planned task by using the candidate base generation model. The refinement node is built based on feedback from the supervisor team to adjust the specific quality issues or defects. 

\textbf{Trial.} Different tools that can achieve the same function are treated as sibling candidate generation nodes under the same parent node. The generation model with the highest priority (selected in the second stage of our pipeline) is placed at the leftmost branch, indicating it will be tried first.

\textbf{Evaluation.} The output of each generation node is evaluated by the supervisor team along three dimensions: (1) Quality - clarity, fidelity, and absence of noise or artifacts; (2) Alignment - consistency of the audio content with the prompt and the generation plan; (3) Aesthetics - the appeal of the audio’s style and its appropriateness to the scene. 

\textbf{Optimization.} If the output meets the evaluation criteria, no further processing is needed; otherwise, an iterative optimization process is triggered. This process has two possible paths: 

\begin{itemize}[leftmargin=2em]
\item \textbf{If the audio has fixable issues} (\textit{e.g.}, an overly long silence at the beginning, slight noise, or unclear sound), the agent creates one or more refinement sub-nodes under the current generation node for targeted post-processing. These refinement nodes do not regenerate content but enhance specific aspect of the audio based on the supervisor's suggestions. 
\item \textbf{If the generation result deviates significantly from the prompt or fails to meet core requirements}, a new sibling generation node is created to retry. The retry process involves prompt adjustments (\textit{e.g.}, refining the text description or adding constraints) or model switching (\textit{e.g.}, selecting another candidate model for generation). This new sibling generation node shares the same parent as the original one but represents a fresh attempt. 
\end{itemize}

By exploring different paths during generation, our module retains flexibility: if one branch fails to produce a satisfactory result, the system can backtrack and try an alternate path. The use of refinement nodes and retry sibling generation nodes embodies the trial-and-error paradigm, which facilitate a continuous loop of evaluation, correction, and iterative improvement of the output.

To improve computational efficiency, we incorporate an early stopping strategy by limiting the depth of the generation tree and the number of retry branches. If the requirements still cannot be met within these limits, the system will return the best result among all explored branches. This mechanism prevents the process from falling into an infinite loop on particularly difficult tasks. However, in practice, trial-and-error iterative optimization typically produces at least one acceptable result before these limits are reached.

\section{Experiment}

\begin{table*}[thb]
\centering
\caption{Comparison with SOTA methods in case of Identical-input, Unconstrained-output Audio Generation. Best performances are highlighted in bold, while second-best are underlined.}% $\uparrow$ & WER $\downarrow$
% \vspace{-0.9em}
\label{SOTA_compare}
\resizebox{\textwidth}{!}{
    \begin{tabular}{c|c|ccccc|cccc}
    \toprule
    \multirow{2}{*}{\shortstack{Input \\ Modalities}}&\multirow{2}{*}{Methods} &\multicolumn{5}{c|}{Objective Metrics} &\multicolumn{4}{c}{Subjective Metrics}\\
    \cmidrule(r){3-7} \cmidrule(r){8-11} 
    && PAM $\uparrow$& PQ $\uparrow$ & PC $\uparrow$ & CE $\uparrow$ & CU $\uparrow$ & MOS-Q $\uparrow$ & MOS-Acc. $\uparrow$ & MOS-Align. $\uparrow$ & MOS-Aesth. $\uparrow$\\
    \midrule
    \multirow{4}{*}{Video}&MMAudio~\cite{cheng2024taming}& \textbf{0.627} & \textbf{6.283} & 2.880 & \underline{3.730} & \underline{5.957} & $\underline{3.15}\pm1.14$ & $\underline{2.90}\pm1.10$ & $\underline{2.62}\pm1.28$ & $\underline{2.62}\pm0.94$ \\
    % &FoleyCrafter~\cite{zhang2024foleycrafter}$^{\ast}$& \underline{0.568} & \underline{6.255} & 2.891 & 3.420 & 5.928 & $2.67\pm1.42$ & $2.44\pm1.35$ & $2.22\pm1.25$ & $2.33\pm0.97$ \\
    % &VidMuse~\cite{tian2024vidmuse}$^{\ast}$& 0.931 & 7.634 & 5.701 & 7.470 & 7.809& \\
    &Seeing and Hearing~\cite{xing2024seeing}& 0.310 & 5.082 & 3.005 & 2.890 & 4.552 & $2.00\pm1.07$ & $2.15\pm1.14$ & $1.85\pm1.13$ & $1.77\pm0.83$ \\
    &AudioX~\cite{tian2025audiox} & 0.459 & 5.742 & \underline{3.018} & 3.502 & 5.339 & $2.23\pm1.27$ & $2.31\pm1.42$ & $2.54\pm1.74$ & $1.92\pm1.14$ \\
    &Ours& \underline{0.551} & \underline{6.212} & \textbf{3.990} & \textbf{4.507} & \textbf{6.106} & $\textbf{3.85}\pm1.05$ & $\textbf{3.54}\pm0.94$ & $\textbf{3.92}\pm0.91$ & $\textbf{3.69}\pm1.15$\\
    \midrule
    \multirow{6}{*}{Text}&Auffusion~\cite{xue2024auffusion}& 0.773 & 5.105 & 3.710 & 5.117 & 4.908 & $1.83\pm0.75$ & $1.80\pm0.84$ & $1.71\pm0.49$ & $1.75\pm 0.96$ \\
    &MMAudio~\cite{cheng2024taming}& \textbf{0.864} & 6.370 & 3.045 & 4.491 & \textbf{6.540} & $\underline{3.25}\pm0.50$ & $\underline{2.50}\pm0.58$ & $\underline{2.25}\pm0.50$ & $\underline{3.25}\pm0.50$ \\
    &AudioX~\cite{tian2025audiox}&0.720 &\underline{6.467}	&3.546 &5.148 &6.071& $2.76\pm1.16$ &$1.76\pm0.59$ &$2.07\pm0.64$ &$2.38\pm0.96$\\
    &Make-An-Audio~\cite{huang2023make} & 0.158 & 6.174 & 3.505 & 3.772 & 5.999 & $1.60\pm0.55$ & $1.40\pm0.55$ & $1.25\pm0.50$ & $1.50\pm0.58$ \\
    &Tango 2~\cite{majumder2024tango}& 0.491 & 6.090 & \underline{4.302} & \textbf{5.799} & 6.059 & $1.75\pm0.50$ & $1.67\pm0.52$ & $1.25\pm0.50$ & $1.83\pm0.75$ \\
    &Ours& \underline{0.799} & \textbf{6.644} & \textbf{5.503} & \underline{5.289} & \underline{6.170} & $\textbf{4.70}\pm0.45$ & $\textbf{4.83}\pm0.41$ & $\textbf{4.80}\pm0.40$ & $\textbf{4.86}\pm0.38$ \\
    \midrule
    \multirow{3}{*}{Image}&See2Sound~\cite{dagli2024see}& 0.375 & \underline{6.161} & \textbf{4.783} & \underline{4.175} & \underline{5.091} & $\underline{2.50}\pm0.83$ & $\underline{1.73}\pm1.17$ & $\underline{1.83}\pm1.60$ & $\underline{2.00}\pm1.15$ \\
    &SSV2A~\cite{guo2024gotta} & \underline{0.464} & 5.035 & 3.437 & 4.108 & 4.403 & $1.33\pm0.52$ & $1.29\pm0.76$ & $1.14\pm0.38$ & $1.17\pm0.41$ \\
    &Ours& \textbf{0.699} & \textbf{7.043} & \underline{4.152} & \textbf{5.567} & \textbf{7.030} & $\textbf{4.42}\pm0.78$ & $\textbf{4.35}\pm0.74$ & $\textbf{4.67}\pm0.51$ & $\textbf{4.33}\pm0.81$ \\
    \midrule
    \multirow{3}{*}{\shortstack{Combination \\ (Video + Text)}}&MMAudio~\cite{cheng2024taming}& \textbf{0.782} & \underline{6.270} & \underline{4.934} & \underline{5.387} & \textbf{5.875} & $\underline{2.63}\pm1.30$ & $\underline{1.50}\pm0.75$ & $1.25\pm0.46$ & $\underline{2.25}\pm1.49$ \\
    % &FoleyCrafter~\cite{zhang2024foleycrafter}$^{\ast}$& 0.391 & 6.073 & 4.448 & 3.476 & \underline{5.537} & $1.50\pm0.54$ & $1.28\pm0.48$ & $1.13\pm0.35$ & $1.33\pm0.52$ \\
    &AudioX~\cite{tian2025audiox}& 0.467 & 6.015 & 3.975 & 4.788 & 5.245 & $1.78\pm0.67$ & $1.33\pm0.50$ & $\underline{1.44}\pm0.73$ & $1.22\pm0.67$ \\
    &Ours& \underline{0.650} & \textbf{6.451} & \textbf{5.619} & \textbf{5.422} & \underline{5.507} & $\textbf{4.88}\pm0.33$ & $\textbf{4.75}\pm0.46$ & $\textbf{4.78}\pm0.44$ & $\textbf{4.63}\pm0.51$ \\
    \bottomrule
    \end{tabular}
}
\end{table*}

\subsection{Dataset}
We introduce the \textbf{M}ulti-type \textbf{A}udio synthesis \textbf{Bench}mark (\textbf{MA-Bench}), comprising 198 professionally curated videos sourced from three main origins: (1) online platforms (YouTube, Bilibili); (2) video material websites; and (3) synthetic generators (Sora~\cite{brooks2024video}). 
Importantly, every video in the dataset features multi-type audio tracks, including diverse sound effects, background music, and human speech or singing. To ensure diversity, MABench spans a wide range of video categories. Figure \ref{fig:dataset} (a) quantifies the distribution across 7 video-level categories and Figure \ref{fig:dataset} (b) shows video frames and its corresponding annotations from three data sources.

\subsection{Implementation Details}
We employ the Qwen2.5-VL-72B-Instruct~\cite{bai2025qwen2} model as our MLLM agent in the generation team, which can effectively interpret both image and video inputs. For the supervisor team, we also rely on Qwen2.5-VL-72B-Instruct in stage 1 and stage 2 to monitor and correct each step. In stage 3, we utilize Audio-Reasoner~\cite{xie2025audio}, a large audio language model supporting advanced reasoning on various audio types, to evaluate, reflect on, and refine the generated single-event audio. Our tool library provides various audio generation methods for different audio types, detailed in Table~\ref{tool_library}.
%We test representative samples from each category in our proposed dataset.

\subsection{Evaluation Metrics}
\subsubsection{Objective Metrics.} Since our outputs include a variety of audio types such as sound effects, speech, music, and songs, traditional metrics that only handle a single type are not appropriate. Therefore, we adopt the Audio Aesthetics Score (AES)~\cite{tjandra2025meta}, which can evaluate the quality of multiple audio types. Specifically, AES breaks down audio aesthetics into four dimensions: Production Quality (PQ), Production Complexity (PC), Content Enjoyment (CE), and Content Usefulness (CU). PQ measures technical quality (\textit{e.g.}, clarity, fidelity, dynamic range, frequency balance, and spatial effects). PC examines the richness and complexity of elements within the audio scene. CE assesses subjective enjoyability, encompassing artistic merit, expressiveness, and listener experience. CU evaluates the practical value of the audio, reflecting its potential reuse in creative projects. In addition, we employ PAM~\cite{deshmukh2024pam}, which provides an overall quality score for speech, music, and environmental sounds, to further assess our results.

\subsubsection{Subjective Metrics.}
We also conduct a human subjective preference test based on the mean opinion score (MOS). Specifically, we evaluate the generated audio from four perspectives: (1) MOS-Quality, measuring clarity, noise level, and listening comfort; (2) MOS-Accuracy, examining whether all audio events in the input are correctly produced without omission or redundancy; (3) MOS-Alignment, assessing the consistency between the audio and the input in both spatiotemporal and semantic dimensions, including coherence with the input’s scene and theme; and (4) MOS-Aesthetic, evaluating whether the audio exhibits aesthetic appeal and emotional impact. We collect ratings from at least 20 participants on a five-point scale: 1 = Bad, 2 = Poor, 3 = Fair, 4 = Good, and 5 = Excellent.

\begin{figure*}[t]
    \centering
    \includegraphics[width=0.78\textwidth]{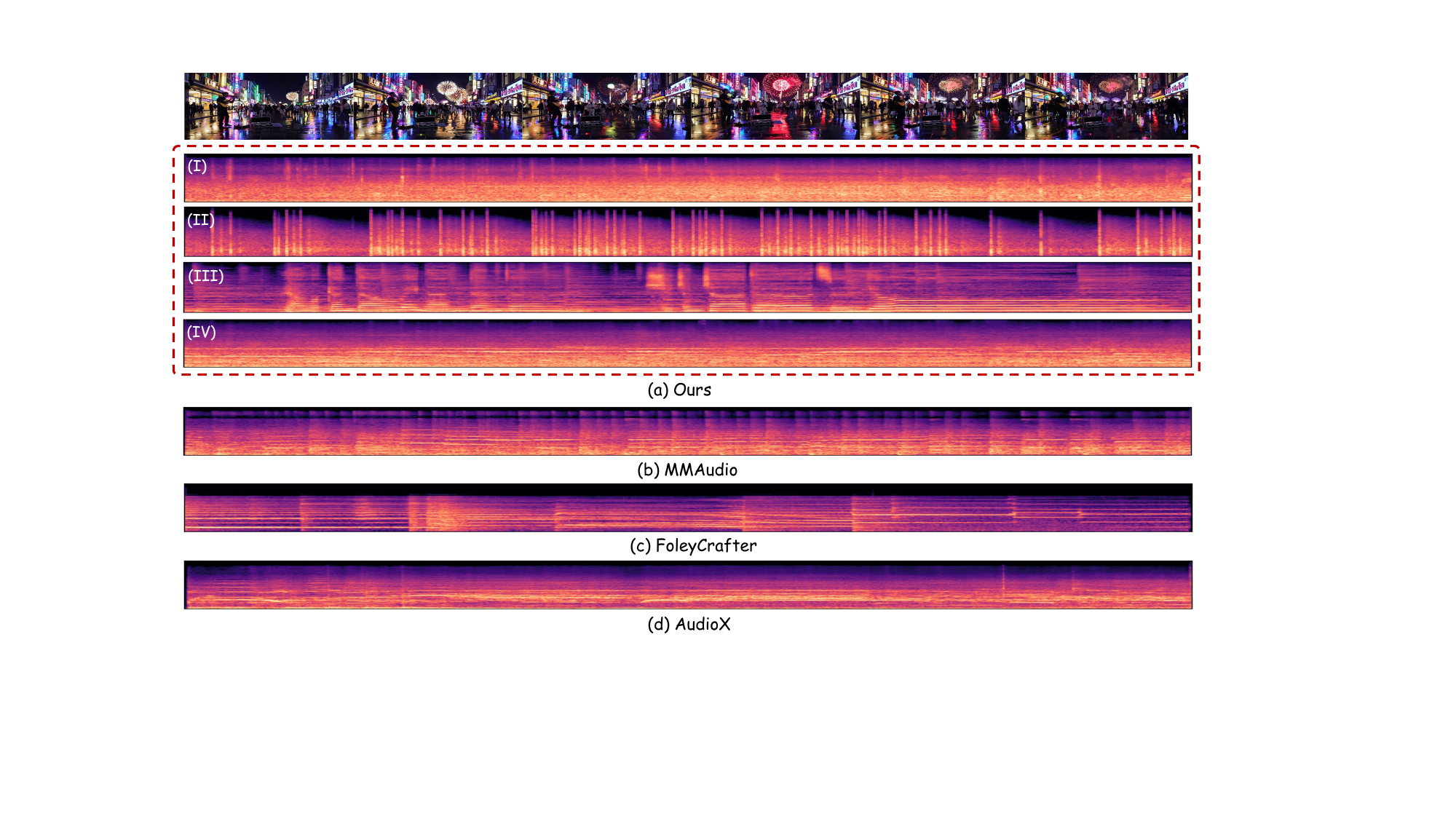}
    \vspace{-0.3em}
    \caption{A visualization example to demonstrate multi-event sound effect generation. The inputs include both video and text: “A bustling commercial street where a street performer sings ‘Chengdu’ (a Chinese song). Pedestrians watch, applaud, and cheer, while fireworks light up the sky, creating a lively scene.” (I)–(IV) are sub-audio events planned by our system: (I) footsteps and cheers of passing pedestrians, (II) fireworks, (III) ambient noise from surrounding shops, and (IV) the song. Our method can generate multiple sound effects based on a video, while other methods can only generate a single sound effect. }
    \vspace{-0.2em}
    \label{fig:melcompare}
\end{figure*}

\subsection{Comparison with SOTA Methods} 
\subsubsection{Identical-input, Unconstrained-output Audio Generation.}
Since no existing approach addresses multimodality-to-multiaudio generation, we conduct a fair comparison by using identical inputs (single input or multi-modal inputs). In this scenario, our method can synthesize multiple audio types (\textit{e.g.}, sound effects, speech, song, or music) from the input, whereas previous approaches generally produce only a single audio category (with the exception of AudioX, which supports two audio type generation). We compare the generation performance between our method and the others. The results are shown in Table~\ref{SOTA_compare}.

Specifically, when using video as input, we compare our method with the latest SOTAs MMAudio~\cite{cheng2024taming}, Seeing and Hearing~\cite{xing2024seeing}, and AudioX~\cite{tian2025audiox}. Among these, MMAudio and Seeing and Hearing generate sound effects aligned with the input video, while AudioX additionally produces music. On MOS-Accuracy and PC, our method achieves scores of 3.54 and 3.99, whereas other methods perform at or below 2.90. This performance gap arises because prior methods often focus on only one event when multiple events occur simultaneously, resulting in incomplete outputs. Moreover, our approach achieves the best results on CE, MOS-Alignment, and MOS-Aesthetic, demonstrating superior spatiotemporal and semantic alignment with the input video, along with enhanced expressiveness and emotional appeal. For FoleyCrafter~\cite{zhang2024foleycrafter}, which can only process videos under 10 seconds, we further select a subset of data within this limit for testing. The results of those experiments are presented in the Appendix.

In the text-input and image-input settings, we compare our method with five text-to-audio generation approaches (MMAudio~\cite{cheng2024taming}, 
Auffusion~\cite{xue2024auffusion}, AudioX~\cite{tian2025audiox}, Make-An-Audio~\cite{huang2023make}, and Tango 2~\cite{majumder2024tango}) and two image-to-audio approaches (See2Sound~\cite{dagli2024see} and SSV2A~\cite{guo2024gotta}), respectively. Our method achieves significantly higher performance on multiple metrics. Existing methods struggle with complex text instructions or images containing multiple audio events, often yielding incomplete or inaccurate generations. Moreover, their outputs commonly exhibit style mismatches relative to the scene in the input image or the theme of the input text. 

We also evaluate inputs containing multiple modalities. Since the image-based module of AudioX is not publicly available, we focus on the video+text-to-audio generation task. We compare two methods: MMAudio and AudioX, each of which accepts both video and text as inputs. Two scenarios are considered: (1) the input text describes information already present in the video, and (2) the input text includes additional information not present in the video. In the first scenario, existing methods can produce semantically appropriate audio but sometimes fail to align correctly in time or space in some scenarios. A possible reason is that these models do not fully capture video content on the visual side, leading them to rely heavily on text inputs and thus causing misalignment. In contrast, our method performs fine-grained audio event analysis and decomposition in stage 1, precisely scheduling each event based on its timestamp in the video, thus improving alignment. In the second scenario, we test cases where the text contains other types of audio (\textit{e.g.}, speech or song). However, existing methods often fail under this setting. We visualize an example in Figure~\ref{fig:melcompare}. When multiple events occur in the video, existing approaches typically capture only a single event, causing other events to be missing from the generated audio. In contrast, our method comprehensively interprets all information in the input, generates each event separately, then mixes them, while employing supervision and self-correction. As a result, all intended audio events are accurately produced.

\begin{table*}[thb]
\centering
\caption{Comparison with SOTA methods for Single-input Single-output Audio Generation. Best performances are highlighted in bold, while second-best are underlined.}
% \vspace{-0.9em}
\label{subtask_compare}
\resizebox{\textwidth}{!}{
    \begin{tabular}{c|c|ccccc|cccc}
    \toprule
    \multirow{2}{*}{Tasks}&\multirow{2}{*}{Methods} &\multicolumn{5}{c|}{Objective Metrics} &\multicolumn{4}{c}{Subjective Metrics}\\
    \cmidrule(r){3-7} \cmidrule(r){8-11} 
    && PAM $\uparrow$& PQ $\uparrow$ & PC $\uparrow$ & CE $\uparrow$ & CU $\uparrow$ & MOS-Q $\uparrow$ & MOS-Acc. $\uparrow$ & MOS-Align. $\uparrow$ & MOS-Aesth. $\uparrow$\\
    \midrule
    \multirow{6}{*}{\shortstack{Audio \\ Effect \\ Generation}}&Tango 2~\cite{majumder2024tango} & 0.708 & 6.169 & \textbf{3.192} & 3.160 & 5.558 & $2.88 \pm 0.64$ & $2.37\pm 1.41$ & $2.13\pm 1.25$ & $2.50\pm 0.93$ \\
    &Make-An-Audio~\cite{huang2023make} & 0.478 & 5.498 & 2.553 & 2.998 & 4.659 & $1.67\pm 0.50$ & $1.22 \pm 0.44 $ & $1.11\pm 0.33$ & $1.22\pm 0.44$ \\
    &Auffusion~\cite{xue2024auffusion} & 0.568 & 6.104 & 2.771 & 2.810 & 5.992 & $3.11 \pm 0.78$ & $2.63 \pm 0.74$ & $2.38 \pm 0.92$ & $2.62 \pm 0.91$ \\
    &MMAudio~\cite{cheng2024taming}& 0.711 & 6.523 & 2.445& 3.018 & \underline{6.059} & $\underline{3.87}\pm1.35$ & $\textbf{3.75}\pm1.58$ & $\textbf{3.87}\pm1.55$ & $\textbf{3.89}\pm1.17$\\
    &AudioX~\cite{tian2025audiox}& \textbf{0.786} & \underline{6.548} & 2.548 & \underline{3.221} & \textbf{6.209} & $3.12\pm1.35$ & $2.63\pm1.60$ & $2.75\pm1.48$ & $2.89\pm1.16$ \\
    &Ours& \underline{0.750} & \textbf{6.595} & \underline{2.931} & \textbf{3.347} & 6.043 & $\textbf{4.11} \pm0.78 $& $\underline{3.44}\pm0.53$ & $\underline{3.67}\pm0.87$ & $\underline{3.78}\pm0.67$ \\
    \midrule
    \multirow{5}{*}{\shortstack{Speech \\ Generation}}&CosyVoice 2~\cite{du2024cosyvoice2} & \underline{0.962} & \underline{7.222} & 1.700 & 5.033 & 5.939 & $\underline{3.33}\pm0.51$ & $\underline{3.83}\pm1.17$ & $\underline{4.16}\pm0.75$ & $\underline{2.83}\pm1.17$ \\
    &F5-TTS~\cite{chen2024f5} & 0.904 & 6.561 & 1.633 & 4.285 & 5.361 & $2.42\pm 1.27$ & $2.71\pm1.89$ & $2.33\pm1.75$ & $1.66\pm0.81$ \\
    &FireRedTTS~\cite{guo2024fireredtts} & 0.953 & 7.096 & \textbf{1.835} & \underline{5.106} & 5.809 & $3.07\pm0.84$ & $3.50\pm1.37$ & $3.42\pm1.27$ & $2.50\pm0.83$ \\
    &VoxInstruct~\cite{zhou2024voxinstruct}& 0.951 & 7.132 & \underline{1.759} & 5.092 & \underline{5.975} & $2.50\pm1.05$ & $2.67\pm 1.63$ & $2.83\pm 1.16$ & $2.07\pm0.83$ \\
    &Ours& \textbf{0.989} & \textbf{7.529} & 1.516 & \textbf{5.549} & \textbf{6.311} & $\textbf{4.83}\pm0.40$ & $\textbf{4.86}\pm0.37$ & $\textbf{4.67}\pm0.51$ & $\textbf{4.50}\pm0.54$   \\
    \midrule
    \multirow{5}{*}{\shortstack{Music \\ Generation}}&InspireMusic~\cite{zhang2025inspiremusic} & 0.912 & 6.325 & 3.489 & \underline{6.784} & 7.376 & $\underline{3.89}\pm0.60$ & $\underline{3.75}\pm0.71$ & $\underline{3.50}\pm1.06$ & $\underline{3.88}\pm0.64$ \\
    &VidMuse~\cite{tian2024vidmuse}&  \underline{0.932} &  \underline{7.449} &  \underline{4.200} & 6.263 & \underline{7.750} & $3.50\pm1.41$ & $2.38\pm1.19$ & $2.13\pm1.35$ & $2.87\pm1.13$ \\
    &MusicGen~\cite{copet2023simple} & 0.409 & 6.569 & 3.715 & 4.718 & 6.576 & $2.88\pm1.81$ & $2.40\pm1.35$ & $2.38\pm1.51$ & $2.44\pm1.59$ \\
    &AudioX~\cite{tian2025audiox}& 0.566 & 6.415 & 3.686 & 4.404 & 5.814 & $3.13\pm1.45$ & $2.75\pm1.04$ & $2.50\pm0.93$ & $3.25\pm1.16$ \\
    &Ours& \textbf{0.974} & \textbf{7.806} & \textbf{5.684} & \textbf{7.527} & \textbf{7.861} & $\textbf{4.88}\pm0.35$ & $\textbf{4.20}\pm0.63$ & $\textbf{4.13}\pm0.64$ & $\textbf{4.25}\pm0.46$ \\
    \midrule
    \multirow{3}{*}{\shortstack{Song \\ Generation}}&DiffRhythm~\cite{ning2025diffrhythm} &  \underline{0.928} & \textbf{8.328} &  \underline{6.418} & \underline{7.464} &  \underline{7.809} & $\underline{3.88}\pm0.64$ & $\underline{3.94}\pm0.52$ & $\underline{3.89}\pm0.53$ & $\underline{4.12}\pm0.35$\\
    &SongGen~\cite{liu2025songgen} & 0.273 & 6.582 & 3.850 & 4.259 & 6.235 & $1.86\pm0.99$ & $1.72\pm0.97$ & $1.75\pm1.04$ & $1.63\pm0.74$ \\
    &Ours& \textbf{0.949} &  \underline{8.212} & \textbf{6.560} & \textbf{7.640} & \textbf{7.812} & $\textbf{4.38}\pm0.74$ & $\textbf{4.00}\pm1.19$ & $\textbf{4.13}\pm1.13$ & $\textbf{4.25}\pm1.03$ \\
    \bottomrule
    \end{tabular}
}
\end{table*}

\subsubsection{Single-input Single-output Audio Generation}
To align with the predominant single-input single-output (SISO) paradigm adopted by existing methods, we constrain our approach to operate under SISO settings for fair comparison with other methods (see Table~\ref{subtask_compare}). For sound effect generation, we benchmark our approach against five other video-to-audio and text-to-audio methods. Our method is slightly lower than MMAudio on a few subjective metrics but outperforms it on most objective metrics. Besides, our method exceeds the performance of all other methods on subjective measures. For speech generation, we compare our system with four recently proposed text-to-speech generation method. Our method achieves better results on multiple metrics. On MOS-Aesthetic, our method scores 4.50, whereas other methods only reach 2.83 or below. A major factor is that, in stage 2, the speech expert analyzes the appropriate state, tone, and emotion based on the input and context, then selects the suitable model and reference prompt. The expert can also add suitable paralinguistic cues, such as laughter, breathing, or emphasis, making the generated speech more expressive and human-like. For the music and song generation sub-tasks, our method achieves the highest scores on all subjective and most objective metrics, thanks to its fine-grained planning across different audio types and adaptive refinement and correction of any issues in the generated audio.

\begin{table}[t]
\centering
\caption{Results of ablation studies on different model components. Best performances are highlighted in bold, while second-best are underlined. SR and CR denote self-refinement and collaborative refinement, respectively.}
\resizebox{\linewidth}{!}{
    \begin{tabular}{c|ccccc}
    \toprule
    Module& PAM $\uparrow$& PQ $\uparrow$ & PC $\uparrow$ & CE $\uparrow$ & CU $\uparrow$ \\
    \midrule
    AudioGenie (Ours)& \textbf{0.966} & \textbf{7.919} & \underline{4.398} & \textbf{6.643} & \textbf{7.167} \\
    w/o supervisor team & 0.888 & 7.547 & 4.111 & 6.190 & 6.850  \\
    w/o domain experts& 0.916 & 7.566 & 4.359 & 6.199 & 6.760\\
    w/o intra-expert SR& 0.941 & 7.849 & 4.388 & 6.509 & \underline{7.008} \\
    w/o inter-expert CR& 0.938 & \underline{7.795} & \textbf{4.445} & \underline{6.540} & 6.953\\
    w/o ToT self-correction& \underline{0.955} & 7.454 & 4.392 & 5.965 & 6.550\\
    \bottomrule
    \end{tabular}
}
\label{table_ablation}
\vspace{-0.3cm}
\end{table}

\subsection{Ablation Studies}
We further examine the importance of different modules through the following ablation settings: (1) w/o supervisor team: remove the supervisor team in the framework, leaving only the generation team; (2) w/o domain-specific experts: replace the domain-specific experts in stage 2 with a single general expert, who handles script refinement and model selection for all audio types; (3) w/o intra-expert self-refinement: disable this mechanism; (4) w/o inter-expert collaborative refinement: disable this mechanism; and (5) w/o Tree-of-thought self-correction: disable trial-and-error iterative refinement in stage 3. The results are shown in Table~\ref{table_ablation}.

Removing the supervisor team causes a noticeable drop in multiple evaluation metrics. This is because step-by-step supervision helps detect deviations or omissions at each stage of planning or generation, preventing the accumulation of errors. Domain-specific experts draw on specialized knowledge for each audio type, enabling more fine-grained and accurate planning of content and style, which substantially improves audio quality and aesthetics. This step forms the foundation for subsequent generation and determines the upper bound of the final performance, whereas a single general expert often struggles with complex tasks requiring diverse expertise. Intra-expert self-refinement and inter-expert collaborative refinement allow agents to refine their outputs through introspection and teamwork, leveraging their strengths while mitigating individual limitations. Notably, the tree-of-thought self-correction module plays a key role in improving the overall quality and aesthetic appeal of generated speech. Single-pass generation methods inevitably encounter errors, but by constructing a tree structure for each audio event and iteratively refining suboptimal outputs, we gain better control and reliability over the final generation.

\section{Conclusion}
In this work, we propose a novel training-free multi-agent system, AudioGenie, which supports multimodal input and multiple types of audio outputs. We introduce a dual-layer coordinated structure comprising a generation team and a supervisor team, which handle the planning and execution of generation, as well as the supervision and correction of each step, respectively. For the generation team, a fine-grained task decomposition and an adaptive MoE collaborative entity are designed for detailed multimodal input understanding and dynamic model selection. We also propose a trial-and-error iterative refinement module that constructs a tree-of-thought for each audio event, enabling continuous improvements by evaluating and revising intermediate outcomes. In addition, we build MA-Bench, the first benchmark for MM2MA generation, which includes 198 annotated videos with multiple audio types. Experiments demonstrate that our AudioGenie achieves SOTA or comparable performance across several metrics. Qualitative assessments and user studies further confirm the effectiveness of our model in terms of quality, accuracy, alignment, and aesthetic.

Future work will focus on: (1) Establishing automatic MM2MA evaluation metrics that comprehensively account for output quality, accuracy, alignment, and aesthetics. (2) Investigating real-time adaptive strategies that incorporate user feedback, allowing the system to refine audio generation on the fly and facilitate interactive multimedia applications.

\begin{acks}
This work was supported by the National Natural Science Foundation of China (No. 62471420), GuangDong Basic and Applied Basic Research Foundation (2025A1515012296), and CCF-Tencent Rhino-Bird Open Research Fund.
\end{acks}

\bibliographystyle{ACM-Reference-Format}
\bibliography{ref}


% \clearpage
\appendix

% \begin{teaser}
\begin{table*}[ht]
\centering
\caption{Comparison with SOTA methods. For a fair comparison with FoleyCrafter, we select videos shorter than 10 seconds. Best performances are highlighted in bold, while second-best are underlined.}
% \vspace{-0.9em}
\label{add_SOTA_compare}
\resizebox{\textwidth}{!}{
\begin{tabular}{c|c|ccccc|cccc}
    \toprule
    \multirow{2}{*}{\shortstack{Input \\ Modalities}}&\multirow{2}{*}{Methods} &\multicolumn{5}{c|}{Objective Metrics} &\multicolumn{4}{c}{Subjective Metrics}\\
    \cmidrule(r){3-7} \cmidrule(r){8-11} 
    && PAM $\uparrow$& PQ $\uparrow$ & PC $\uparrow$ & CE $\uparrow$ & CU $\uparrow$ & MOS-Q $\uparrow$ & MOS-Acc. $\uparrow$ & MOS-Align. $\uparrow$ & MOS-Aesth. $\uparrow$\\
    \midrule
    \multirow{5}{*}{Video}&MMAudio~\cite{cheng2024taming}& \textbf{0.645} & \textbf{6.519} & 2.930 & \underline{3.881} & \underline{6.224} & $\underline{3.09}\pm1.33$ & $\underline{2.70}\pm1.20$ & $\underline{2.36}\pm1.34$ & $2.30\pm1.06$ \\
    &FoleyCrafter~\cite{zhang2024foleycrafter}& \underline{0.568} & 6.255 & 2.891 & 3.420 & 5.928 & $2.67\pm1.42$ & $2.44\pm1.35$ & $2.22\pm1.25$ & $\underline{2.33}\pm0.97$ \\
    % &VidMuse~\cite{tian2024vidmuse}$^{\ast}$& 0.931 & 7.634 & 5.701 & 7.470 & 7.809& \\
    &Seeing and Hearing~\cite{xing2024seeing}& 0.337 & 5.142 & \underline{3.034} & 2.936 & 4.760 & $2.10\pm1.20$ & $2.30\pm1.25$ & $2.09\pm1.25$ & $1.80\pm0.92$ \\
    &AudioX~\cite{tian2025audiox} & 0.437 & 5.656 & 2.915 & 3.340 & 5.194 & $2.37\pm1.34$ & $2.19\pm1.55$ & $2.27\pm1.75$ & $1.91\pm1.10$  \\
    &Ours& 0.558 & \underline{6.448} & \textbf{4.092} & \textbf{4.730} & \textbf{6.239} & $\textbf{3.90}\pm0.99$ & $\textbf{3.50}\pm0.85$ & $\textbf{3.88}\pm0.84$ & $\textbf{3.70}\pm0.94$ \\
    \midrule
    \multirow{4}{*}{\shortstack{Combination \\ (Video + Text)}}&MMAudio~\cite{cheng2024taming}& \textbf{0.841} & 5.935 & \underline{4.573} & \underline{5.098} & \textbf{5.797} & $\underline{2.50}\pm1.37$ & $\underline{1.67}\pm0.81$ & $1.33\pm0.51$ & $\underline{2.17}\pm1.47$\\
    &FoleyCrafter~\cite{zhang2024foleycrafter}& 0.391 & \underline{6.073} & 4.448 & 3.476 & \underline{5.537} & $1.50\pm0.54$ & $1.28\pm0.48$ & $1.13\pm0.35$ & $1.33\pm0.52$ \\
    &AudioX~\cite{tian2025audiox}& 0.425 & 5.806 & 3.859 & 4.563 & 4.988 & $1.71\pm0.75$ & $1.43\pm0.53$ & $\underline{1.57}\pm0.78$ & $1.28\pm0.75$ \\
    &Ours& \underline{0.651} & \textbf{6.465} & \textbf{5.261} & \textbf{5.150} & 5.223 & $\textbf{4.86}\pm0.38$ & $\textbf{4.67}\pm0.51$ & $\textbf{4.71}\pm0.48$ & $\textbf{4.50}\pm0.55$ \\
    \bottomrule
\end{tabular}
}
\end{table*}
% \end{teaser}

\begin{table*}[ht]
\centering
\caption{Comparison of AudioGenie and SOTA methods on VGGSound and V2M. Best performances are highlighted in bold, while second-best are underlined.}
\label{tab:public_data}
\begin{tabular}{c|ccccc|cccc}
    \toprule
    \multirow{2}{*}{Methods} &\multicolumn{5}{c|}{Objective Metrics} &\multicolumn{4}{c}{Subjective Metrics}\\
    \cmidrule(r){2-6} \cmidrule(r){7-10} 
    & PAM $\uparrow$ & PQ $\uparrow$ & PC $\uparrow$ & CE $\uparrow$ & CU $\uparrow$ & MOS-Q $\uparrow$ & MOS-Acc. $\uparrow$ & MOS-Align. $\uparrow$ & MOS-Aesth. $\uparrow$ \\
    \midrule
    MMAudio~\cite{cheng2024taming} & \textbf{0.632} & \textbf{6.226} & \underline{3.334} & \underline{3.158} & 4.861 & \underline{3.88} $\pm$ 0.60 & \underline{4.04} $\pm$ 0.81 & \underline{4.12} $\pm$ 0.79 & \underline{3.80} $\pm$ 0.77 \\
    Seeing and Hearing~\cite{xing2024seeing} & 0.329 & 5.035 & 2.985 & 2.869 & 4.339 & 1.98 $\pm$ 0.85 & 1.96 $\pm$ 0.86 & 1.87 $\pm$ 0.83 & 1.79 $\pm$ 0.84 \\
    AudioX~\cite{tian2025audiox} & 0.478 & 5.749 & 2.875 & 2.915 & \textbf{5.086} & 3.06 $\pm$ 1.04 & 3.17 $\pm$ 1.07 & 2.98 $\pm$ 1.14 & 2.87 $\pm$ 0.99 \\
    Ours & \underline{0.617} & \underline{6.060} & \textbf{3.804} & \textbf{3.473} & \underline{5.014} & \textbf{4.10} $\pm$ 0.64 & \textbf{4.29} $\pm$ 0.68 & \textbf{4.19} $\pm$ 0.85 & \textbf{4.15} $\pm$ 0.78 \\
    \bottomrule
\end{tabular}
\end{table*}

\begin{table*}[ht]
\centering
\caption{Performance comparison between AudioGenie (using various MLLMs) and other SOTA methods on the MA-Bench video-input task. Best performances are highlighted in bold, while second-best are underlined.}
\label{tab:diff_mllm}
\begin{tabular}{c|ccccc|cccc}
    \toprule
    \multirow{2}{*}{Methods} &\multicolumn{5}{c|}{Objective Metrics} &\multicolumn{4}{c}{Subjective Metrics}\\
    \cmidrule(r){2-6} \cmidrule(r){7-10} 
    & PAM $\uparrow$ & PQ $\uparrow$ & PC $\uparrow$ & CE $\uparrow$ & CU $\uparrow$ & MOS-Q $\uparrow$ & MOS-Acc. $\uparrow$ & MOS-Align. $\uparrow$ & MOS-Aesth. $\uparrow$\\
    \midrule
    MMAudio~\cite{cheng2024taming} & \textbf{0.584} & \underline{6.341} & 2.956 & 3.406 & 5.387 & $3.28 \pm 0.78$ & $3.14 \pm 1.01$ & $3.04 \pm 0.81$ & $2.76 \pm 0.70$ \\
    Seeing and Hearing~\cite{xing2024seeing} & 0.186 & 5.274 & 2.143 & 2.360 & 3.836 & $2.38 \pm 1.36$ & $2.28 \pm 1.23$ & $2.33 \pm 1.15$ & $2.23 \pm 1.17$ \\
    AudioX~\cite{tian2025audiox} & 0.310 & 6.138 & 2.833 & 3.371 & 5.622 & $2.42 \pm 1.07$ & $2.14 \pm 0.96$ & $2.28 \pm 1.01$ & $2.14 \pm 0.91$ \\
    \midrule
    Ours (Qwen)       & \underline{0.578} & \textbf{6.546} & 4.215 & \textbf{4.196} & \underline{5.863} & $\underline{3.76} \pm 0.83$ & $\underline{3.71} \pm 0.78$ & $\textbf{3.81} \pm 0.92$ & $3.43 \pm 1.02$ \\
    Ours (GPT-4o)     & 0.544 & 6.270 & \underline{4.278} & 3.936 & 5.352 & $3.67 \pm 0.79$ & $3.48 \pm 0.75$ & $3.38 \pm 0.92$ & $\underline{3.62} \pm 0.74$ \\
    Ours (Gemini)     & 0.519 & 6.272 & \textbf{4.430} & \underline{4.126} & \textbf{6.035} & $\textbf{3.91} \pm 0.70$ & $\textbf{3.76} \pm 0.70$ & $\underline{3.57} \pm 0.87$ & $\textbf{3.63} \pm 0.59$ \\
    \bottomrule
\end{tabular}
\end{table*}

\section{Additional SOTA Comparisons on Short Videos}
Given that FoleyCrafter~\cite{zhang2024foleycrafter} only supports audio under 10 seconds, we further evaluate different methods on short videos for a more comprehensive comparison, considering both video-only and video+text inputs. The results are shown in Table~\ref{add_SOTA_compare}. Specifically, for video-only inputs, our approach is slightly lower on the objective metrics PAM and PQ compared to other methods, but it achieves the best performance on all subjective metrics, and outperforms others on PC, CE, and CU. In the video+text input scenario (with the same experimental setting in Table 3), our method attains the best results in terms of audio quality, completeness, alignment, and aesthetics.

\section{Generalization of results to public datasets}
We further evaluate AudioGenie on two widely used public datasets, VGGSound~\cite{chen2020vggsound} and V2M~\cite{tian2024vidmuse}, with quantitative results in Table~\ref{tab:public_data}. For single-event generation, our method achieves performance comparable to SOTA V2A models. However, in complex, multi-event scenarios, existing methods often collapse to a single dominant event, leading to incomplete audio generation. In contrast, our method successfully captures multiple events, ensuring better spatio-temporal and semantic alignment with the input video. Furthermore, the audio generated by our framework exhibits superior quality and realism.

\section{Results on Different MLLMs}
To show the scalability of our framework, we replace Qwen2.5-VL-72B-Instruct with GPT-4o~\cite{openai2024gpt4o} and Gemini2.5-Pro~\cite{google2025gemini2_5_pro} and tested on the V2A task. As detailed in Table~\ref{tab:public_data}, our method outperforms existing SOTA V2A models when using any of these MLLMs as the agent. We also observe that different MLLMs exhibit distinct advantages on various metrics, indicating that our framework is not reliant on a specific MLLM.

\end{document}